\documentclass[twocolumn]{aastex6}

\usepackage{natbib}
\newcommand{\keyw}[1]{\textcolor{gray}{#1}}

\begin{document}

\title{Ringed Substructure and a Gap at 1 AU in the Nearest Protoplanetary Disk}
\author{
Sean M.~Andrews\altaffilmark{1}, 
David J.~Wilner\altaffilmark{1},
Zhaohuan Zhu\altaffilmark{2},
Tilman Birnstiel\altaffilmark{3},
John M.~Carpenter\altaffilmark{4}, \\
Laura M.~P{\'e}rez\altaffilmark{5}, 
Xue-Ning Bai\altaffilmark{1},
Karin I.~{\"O}berg\altaffilmark{1},
A.~Meredith Hughes\altaffilmark{6},
Andrea Isella\altaffilmark{7},
\& Luca Ricci\altaffilmark{1}
}
\altaffiltext{1}{Harvard-Smithsonian Center for Astrophysics, 60 Garden Street, Cambridge, MA 02138, USA; sandrews@cfa.harvard.edu}
\altaffiltext{2}{Department of Astrophysical Sciences, 4 Ivy Lane, Peyton Hall, Princeton University, Princeton, NJ 08544, USA}
\altaffiltext{3}{Max-Planck-Institut f{\"u}r Astronomie, K{\"o}nigstuhl 17, 69117 Heidelberg, Germany}
\altaffiltext{4}{Joint ALMA Observatory (JAO), Alonso de Cordova 3107 Vitacura -Santiago de Chile}
\altaffiltext{5}{Max-Planck-Institut f{\"u}r Radioastronomie, Auf dem H{\"u}gel 69, 53121 Bonn, Germany}
\altaffiltext{6}{Department of Astronomy, Wesleyan University, Van Vleck Observatory, 96 Foss Hill Drive, Middletown, CT 06457, USA}
\altaffiltext{7}{Department of Physics and Astronomy, Rice University, 6100 Main Street, Houston, TX, 77005, USA}

\begin{abstract}
We present long-baseline Atacama Large Millimeter/submillimeter Array (ALMA) observations of the 870\,$\mu$m continuum emission from the nearest gas-rich protoplanetary disk, around TW Hya, that trace millimeter-sized particles down to spatial scales as small as 1 AU (20\,mas).  These data reveal a series of concentric ring-shaped substructures in the form of bright zones and narrow dark annuli (1--6\,AU) with modest contrasts (5--30\%).  We associate these features with concentrations of solids that have had their inward radial drift slowed or stopped, presumably at local gas pressure maxima.  No significant non-axisymmetric structures are detected.  Some of the observed features occur near temperatures that may be associated with the condensation fronts of major volatile species, but the relatively small brightness contrasts may also be a consequence of magnetized disk evolution (the so-called zonal flows).  Other features, particularly a narrow dark annulus located only 1 AU from the star, could indicate interactions between the disk and young planets.  These data signal that ordered substructures on $\sim$AU scales can be common, fundamental factors in disk evolution, and that high resolution microwave imaging can help characterize them during the epoch of planet formation.   
\end{abstract}

\section{Introduction}

The disks around young stars are the formation sites of planetary systems.  However, the smooth, monotonic radial distributions of gas temperatures and densities assumed in most theoretical models of planetary formation create a fundamental dilemma.  The millimeter-sized particles needed to assemble larger planetesimals \citep{ormel10,lambrechts12} experience aerodynamic drag with the gas that results in their rapid migration toward the host star \citep{weidenschilling77,takeuchi02}.  Yet, this predicted depletion of solids is not commensurate with observations that routinely detect microwave continuum emission from such particles extending over a large range of disk radii, out to tens or hundreds of AU (see reviews by \citealt{williams11} or \citealt{andrews15}).  The hypothesized solution to this conflict invokes substructure in the form of local gas pressure maxima, which slow or stop the migration of these particles and concentrate the solid densities to levels that might trigger efficient planetesimal growth \citep[e.g.,][]{whipple72,pinilla12a}.     

The young solar analog TW Hya is an especially interesting target to characterize disk substructures, for three primary reasons.  First, it is the closest gas-rich disk to Earth \citep[$54\pm6$\,pc;][]{vanleeuwen07}, providing unique access to its properties at incomparably fine levels of detail.  Second, it is the benchmark case study for an evolved population of disk solids, with a radially concentrated population of larger ($\gtrsim$cm) particles \citep{wilner05,menu14} and a sharp radial decrease in the solids-to-gas mass ratio \citep{andrews12,birnstiel14} that announce substantial growth and inward migration.  The predicted depletion of microwave emission due to radial drift \citep{takeuchi05} should be especially prominent at the relatively advanced age of TW Hya \citep[$\sim$10 Myr; e.g.,] []{weinberger13}; but since that is not what is observed, the case for relaxing the assumption of a globally negative pressure gradient is bolstered.  And third, there is already tantalizing evidence for substructure in this disk, including a central depletion \citep{calvet02,hughes07} and tentative signatures of ``gaps" or ``breaks" in the infrared scattered light emission \citep{akiyama15,rapson15,debes13,debes16}. 

In this Letter, we present and analyze observations that shed new light on the substructure in the TW Hya disk.  We have used the long baselines of ALMA to measure the 870\,$\mu$m continuum emission from this disk at an unprecedented spatial resolution of $\sim$1\,AU.  Section~\ref{sec:data} presents these observations, Section~\ref{sec:analysis} describes a broadbrush analysis of the continuum data, and Section~\ref{sec:discussion} considers potential interpretations of the results in the contexts of disk evolution and planet formation.

\section{Observations and Data Calibration} \label{sec:data}

TW Hya was observed by ALMA on 2015 November 23, November 30, and December 1.  The array included 36, 31, and 34 antennas, respectively, configured to span baseline lengths from 20\,m to 14\,km.  The correlator processed four spectral windows centered at 344.5, 345.8, 355.1, and 357.1\,GHz with bandwidths of 1875, 469, 1875, and 1875\,MHz, respectively.  The observations cycled between the target and J1103-3251 with a 1 minute cadence.  Additional visits to J1107-3043 were made every 15 minutes.  J1037-2934, J1058+0133, and J1107-4449 were briefly observed as calibrators.  The precipitable water vapor (PWV) levels were $\sim$1.0\,mm on November 23 and 0.7\,mm on the latter two executions.  The total on-target integration time was $\sim$2\,hours.

These raw data were calibrated by NRAO staff.  After applying phase corrections from water vapor radiometer measurements, the data were time-averaged into 2 s integrations and flagged for problematic antennas and times.  The bandpass response of each spectral window was calibrated using the observations of J1058+0133.  The amplitude scale was determined from J1037-2934 and J1107-4449.  The complex gain response of the system was calibrated using the frequent observations of J1103-3251.  Although images generated from these data are relatively free of artifacts and recover the integrated flux density of the target (1.5\,Jy), folding in additional ALMA observations with a higher density of short antenna spacings improves the image reconstruction.

To that end, we calibrated three archival ALMA observations of TW Hya, from 2012 May 20, 2012 Nov 20, and 2014 Dec 31, using 16, 25, and 34 antennas spanning baselines from 15--375\,m.  The first two observations had four 59\,MHz-wide spectral windows centered at 333.8, 335.4, 345.8, and 347.4 GHz.  The latter had two 235\,MHz windows (at 338.2 and 349.4\,GHz), one 469\,MHz window (at 352.0\,GHz), and one 1875\,MHz window (at 338.4\,GHz).  J1037-2934 was employed as a gain calibrator, and Titan and 3C 279 (May 20), Ceres and J0522-364 (Nov 20), or Ganymede and J0158+0133 (Dec 31) served as flux or bandpass calibrators.  The weather for these observations was excellent, with PWV levels of 0.5--1\,mm.  The combined on-target integration time was 95 minutes.  The basic calibration was as described above.  As a check, we compared the amplitudes from each individual dataset on overlapping spatial frequencies and found exceptional consistency.

The calibrated visibilities from each observation were shifted to account for the proper motion of the target and then combined after excising channels with potential emission from spectral lines.  Some modest improvements were made with a round of phase-only self-calibration.  Continuum images at a mean frequency of 345.9\,GHz (867\,$\mu$m) were generated by Fourier inverting the visibilities, deconvolving with a multi-scale, multi-frequency synthesis version of the {\tt CLEAN} algorithm, and then restoring with a synthesized beam.  All calibration and imaging was performed with the {\tt CASA} package (v4.5.0).     

After some experimentation, we settled on an analysis of two images made from the same composite dataset.  The first used a Briggs weighting (with a robust parameter of 0) to provide a $24\times18$\,mas synthesized beam (at P.A.=78\degr).  While this provides enhanced resolution, it comes at the cost of a dirty beam with $\sim$20\%\ sidelobes (due to the sparse coverage at long baselines) that degrades the image quality.  A second image was made with a robust parameter of 0.5 and an elliptical taper to create a circular 30\,mas beam with negligible sidelobes.  Both images are consistent (within the resolution differences) and have RMS noise levels around 35\,$\mu$Jy beam$^{-1}$.

\section{Results} \label{sec:analysis}

\begin{figure}
\includegraphics[width=\linewidth]{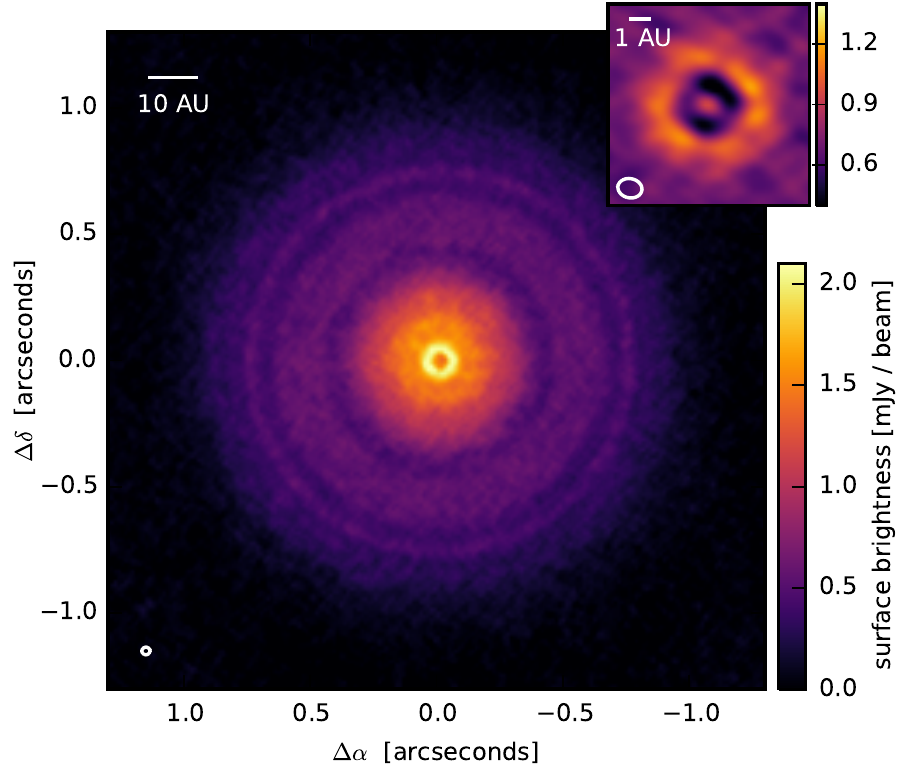}
\figcaption{A synthesized image of the 870\,$\mu$m continuum emission from the TW Hya disk with a 30\,mas FWHM (1.6\,AU) circular beam.  The RMS noise level is $\sim$35\,$\mu$Jy beam$^{-1}$.  The inset shows a 0.2\arcsec-wide (10.8\,AU) zoom using an image with finer resolution ($24 \times 18$\,mas, or $1.3 \times 1.0$\,AU, FWHM beam).
\label{fig:image}}
\end{figure}

\begin{figure}
\includegraphics[width=\linewidth]{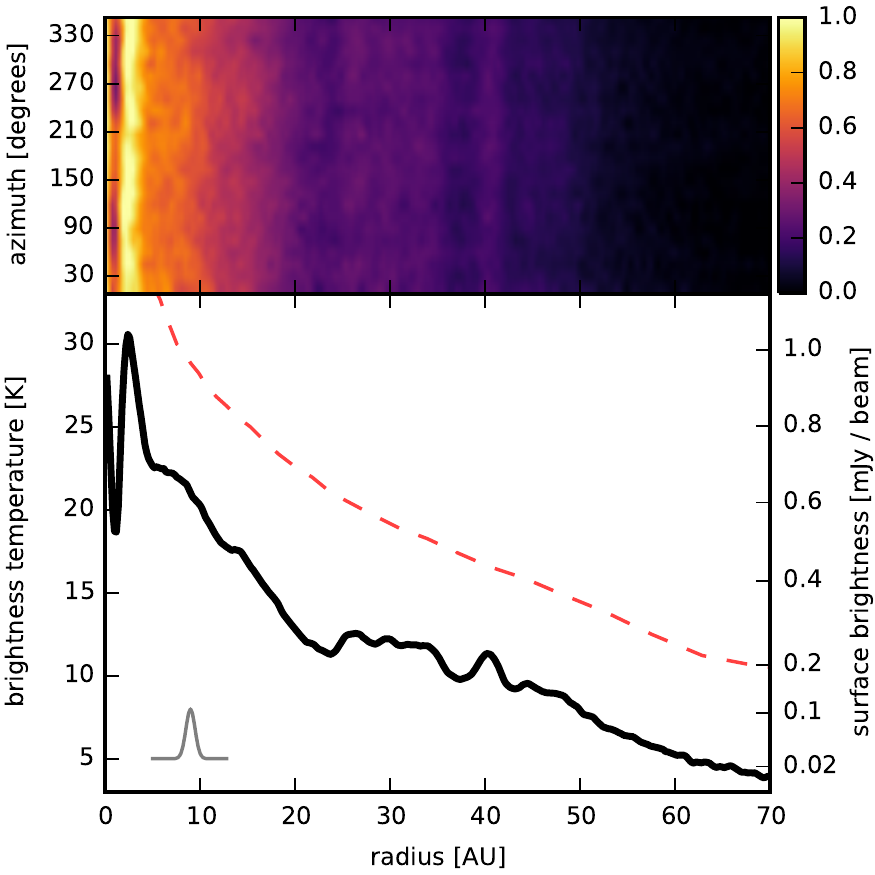}
\figcaption{({\it top}) The high resolution ($24\times18$\,mas beam) synthesized image described in Sect.~\ref{sec:data}, deprojected into a map in polar coordinates to more easily view the disk substructure.  ({\it bottom}) The azimuthally-averaged radial surface brightness profile.  For reference, the dashed red curve shows the midplane temperature profile derived from a representative model disk.  The gray curve in the bottom left reflects the profile of the synthesized beam.  
\label{fig:raz_rprof}}
\end{figure}

Figure~\ref{fig:image} shows a high resolution map of the 870\,$\mu$m continuum emission from the TW Hya disk, revealing a series of concentric bright and dark rings out to a radial distance of 60\,AU from the host star with a nearly pole-on viewing geometry.  To aid in the visualization of this substructure, Figure~\ref{fig:raz_rprof} shows the image transformed into polar coordinates and azimuthally averaged into a radial surface brightness profile.  

The inner disk includes an unresolved ($<$0.5\,AU in radius) $0.93\pm0.04$\,mJy source  coincident with the stellar position and a bright ring that peaks at 2.4\,AU; between them is a dark annulus centered at 1\,AU.  The bright ring and dark annulus are unresolved ($<$1\,AU across).  Because it is unresolved, the depth of the dark annulus is difficult to determine unambiguously: we find at least a 30\%\ brightness reduction.  

The outer disk shows dark annuli at 22, 37, and 43\,AU separated by more extended bright zones.  We tentatively identify additional dark annuli that have lower contrasts with respect to their neighboring regions at 12, 28, and 31\,AU.  These dark features are narrow, only 1--6\,AU across, and shallow, with depths only 5--20\%\ below the intensities of neighboring regions.  The dark annulus at 22\,AU has also been seen at infrared wavelengths, as a deficit of light reflected off the disk surface \citep{akiyama15,rapson15,debes16}.  \citet{zhang16} also recently interpreted an unresolved feature near this radius in their 0\farcs3 resolution ALMA images as a ``turning point" in the intensity profile.\footnote{\citet{nomura15} identify a similar feature, but the extreme spatial filtering applied in their imaging introduces severe artifacts that masquerade as ``gaps" and muddle the interpretation.}  \citet{debes13} reported an additional annular deficit in the scattered light around 80\,AU, but there is too little microwave continuum emission that far from the star to identify any related features in the ALMA data.  

Based on the data in Figure~\ref{fig:raz_rprof}, comparisons of the radial profile along small ranges of polar angles in the disk plane demonstrate that there are no statistically significant azimuthal variations in the observed emission.  The typical deviations correspond roughly to the RMS noise level; the median fractional deviation is only 7\%.  We fitted ellipses to the prominent bright rings at 2.4 and 40\,AU (the latter between two dark annuli) and the dark annulus at 22\,AU to estimate the projected viewing geometry of the disk.  Each feature was consistent with the standard geometry inferred from molecular line observations \citep[e.g.,][]{qi04,andrews12}: the joint constraints suggest an inclination of $7\pm3$\degr\ and a major axis position angle of $155\pm12$\degr.  

On broad angular scales, the overall continuum brightness distribution is roughly consistent with the broken power-law model of \citet{hogerheijde16}: the surface brightness falls off like $r^{-0.7}$ inside 50\,AU and then drops precipitously, like $r^{-6}$, at larger radii.  A refined description would characterize the emission inside 50\,AU using two different segments, where the linear slope inside 20\,AU is about 6$\times$ steeper than from 20--50\,AU.  The fact that this slope change occurs near the 22\,AU dark annulus is certainly (at least partially) associated with the feature noticed by \citet{zhang16} and \citet{nomura15}: after all, the dark annulus itself would not be detectable at resolutions coarser than $\sim$0\farcs1.  

For some general guidance, we generated a representative (but not in any sense optimized) model disk using the radiative transfer code {\tt RADMC-3D} \citep{dullemond12}.  We followed the same basic assumptions as \citet{andrews12}, with a few distinctions: (1) we employed a broken power-law dust surface density prescription \citep{hogerheijde16}; (2) in the very inner disk, we included a belt of dust that has a Gaussian density profile with a width of 0.4\,AU (to insure it would be unresolved); (3) to mimic the observed dark annulus, the outer disk is truncated at 2.4\,AU (rather than the previously assumed 4\,AU).  With only minor tweaks to the original parameters, this model reproduces well the broadband spectral energy distribution.  It also accounts for the $\sim$1\,mJy point source at the disk center, confirming its likely origin as thermal emission from warm ($\sim$200\,K) dust adjacent to the host star.  However, it still has difficulty reproducing the observed brightness profile in detail.  The midplane temperature profile is shown in the bottom panel of Figure~\ref{fig:raz_rprof}.  While we aim to refine this model in future work, for now it serves as a crude reference for the disk temperatures that will aid in our discussion of the potential mechanisms responsible for the observed substructure.

\section{Discussion} \label{sec:discussion}

Currently only one other disk, around the much younger source HL Tau, is known to exhibit ringed substructure like we find here \citep{brogan15}.  But the apparent similarities are superficial.  The dark annuli in the TW Hya disk are considerably less deep and narrower: if TW Hya were located at the same distance as HL Tau (140\,pc), the same data would at best only be able to (tentatively) identify the 22\,AU feature.  Unlike in the HL Tau disk, the bright zones we observe have only marginal optical depths (given our representative temperature model and others like it).  The slope transition noted at $\sim$20\,AU seems to mark the change to an optically thick inner disk (aside from the 1\,AU dark annulus).  In any case, it is interesting that this type of ordered substructure is observed in these two very different disks, which bracket a factor of $\sim$10 in both their ages and microwave luminosities (the TW Hya disk being older and intrinsically less luminous).  

While the new data presented here corroborate the emerging concept that well-ordered, azimuthally symmetric substructures are prevalent and important forces in disk evolution, they also extend the interesting diversity we have seen in terms of the amplitudes and physical scales on which that substructure is manifested.  Various mechanisms to produce ringed substructure in disks have been proposed.  They can be broadly categorized into magnetic, chemical, or dynamical origins: here we consider each in the context of the TW Hya disk.  

Magnetized disks may exhibit radial pressure variations known as zonal flows \citep{johansen09}, natural outcomes of the turbulence driven by the magneto-rotational instability \citep{balbus91}.  Numerical simulations in a local box show that zonal flow pressure maxima are separated by a few scale heights and can have amplitudes as high as $\sim$30-50\%, depending on the magnetic field configuration \citep[e.g.,][]{simon14,bai14b}.  The scale heights in our representative model are 1--3\,AU at radii of 20--40\,AU.  Both the observed contrasts and separations of the ring-like features beyond $\sim$20\,AU are consistent with this scenario.  While global simulations with realistic disk physics and magnetic fields remain computationally challenging, low-amplitude surface density variations are present in the recent global simulations assuming ideal MHD by \citet{suzuki14}.  The current observational constraints on turbulent linewidths in the TW Hya disk \citep{hughes11} are sensitive to much larger radii; it remains unclear if there is sufficient MHD turbulence in the 20--50\,AU range to produce zonal flows consistent with the ALMA image.  However, \citet{bai15} found that even a thin turbulent layer at the disk surface (owing to far-ultraviolet ionization; e.g., \citealt{perez-becker11}) appears sufficient to drive such flows down to the midplane, even if it is largely laminar.  Ring and gap features have also been found in simulations that consider the dynamical effects of radial variations in the disk resistivity \citep[e.g.,][]{flock15,lyra15}.\footnote{Although we note that other non-ideal MHD effects were not included in these studies, and that this scenario may not be able to account for multiple ring-like features.}   Overall, low-amplitude radial surface density variations may be a generic consequence of magnetized disk evolution.  
On the other hand, it is less clear whether magnetic mechanisms alone could account for the ring at 2.4\,AU and the dark annulus at 1\,AU, where the more complex disk microphysics (thermodynamics, ionization, and non-ideal MHD effects) is not very well understood.

The chemical explanation is elegant and should be universal.  As migrating solids approach the condensation fronts of major volatile species, they shed their corresponding ices back into the gas phase.  This and any subsequent re-condensation can modify the solid opacities, and thereby the associated continuum emission \citep[e.g.,][]{cuzzi04}.  This sublimation may also make the particles brittle enough that collisions become destructive, leaving small fragments that are better coupled to the gas, and therefore migrate much more slowly.  The net result is a sequence of “traffic jams” that locally enhance the solid densities and would appear as bright continuum rings at a set of specific temperatures \citep{okuzumi16}.  This scenario is different than the magnetic and dynamical mechanisms, in that it is not precipitated by substructure in the gas disk and therefore does not ``trap" particles.  Nevertheless, it is perhaps appealing in the outer disk of TW Hya, where the carbon monoxide (CO) and molecular nitrogen (N$_2$) condensation fronts at temperatures of $\sim$20 and 17\,K lie outside the 22 and 37\,AU dark annuli, respectively, in our reference model.  The former also overlaps with the inner edge of the N$_2$H$^+$ ring found by \citet{qi13} and associated with the CO snowline.  However, it is worth a reminder that temperature models are subject to uncertainties associated with detailed assumptions about the disk structure and grain properties: small changes to the optical depth profile can substantially shift the radii that correspond to the relevant volatile condensation temperatures.

It may be tempting to consider the H$_2$O snowline as a potential cause for the substructure observed at $\sim$1--3\,AU.  Theoretical models predict significant changes in the solid properties (e.g., sticking probabilities, material abundances, and particle sizes) around this particular phase transition \citep{birnstiel10,banzatti15}.  Our representative model places the appropriate midplane condensation temperature, $\sim$150\,K, at radii $<$1\,AU, but it is not unreasonable to assume that alternative surface density or opacity prescriptions could push it further out.  Nevertheless, we do not see a good reason to expect that the H$_2$O snowline itself would result in the substantial density depletion of a wide range of particle sizes ($\mu$m--cm) that would be required to explain both the pronounced dip in the mid-infrared spectrum \citep{calvet02} and the 1\,AU dark annulus in the ALMA image.  Moreover, we would expect a much higher than observed incidence of TW Hya-like infrared spectra (i.e., classical ``transition" disks) if this were the case, since essentially all disks around young solar analogs should have H$_2$O snowlines at similar locations. 

The dynamical alternative instead postulates that the dark annuli are true gaps that have had their densities depleted by interactions with (as yet unseen) planetary-mass companions \citep{lin86,kley12}.  Given the scale heights implied by our representative model temperatures, even relatively low-mass ($\sim$several $M_{\oplus}$) planets can open narrow gaps \citep{zhu13,duffell13,fung14} and trap solid particles \citep{paardekooper06,zhu14,picogna15} to produce features similar to the substructure that we observe here.  Even with such low masses, it seems difficult to explain both the features at 37 and 43 with two planets since they would tend to open a common gap without the bright ring observed to bisect them.  In principle, a single low-mass planet at $\sim$40\,AU could potentially open a `double-gap' feature if the disk is particularly inviscid \citep[e.g.,][]{goodman01,dong11,duffell12}.  Or, interactions with a planet at $\sim$37\,AU might also perturb the distribution of solids out of the disk plane beyond it; the shadowing induced by such surface variations can generate temperature variations that might mimic the 43\,AU feature \citep[e.g.,][]{jang-condell09}.  The 1\,AU dark annulus is an especially compelling case study.  If a more comprehensive modeling of the ALMA data confirms that the observed low brightness contrast indeed corresponds to a density depletion factor like that expected from a young super-Earth, it could serve as a touchstone for modeling the formation of the large population of such planets identified with the {\it Kepler} mission \citep[e.g.,][]{howard13b}.

Regardless of which of these mechanisms are at work, the new ALMA data we have presented suggest that symmetric, well-ordered, substructure is prevalent in the disks around young stars, down to very small physical scales (comparable to or smaller than the local scale height) and with a range of amplitudes.  Such features are the observational hallmarks of the long-speculated solution to the fundamental problem of the fast migration of disk solids that has subverted the standard theory of the planet formation process for decades.

\acknowledgments We thank Ian Czekala for advice on the figures, the NRAO/NAASC staff for their assistance with the data calibration, and Ilse Cleeves and Ruobing Dong for helpful conversations.  This research greatly benefitted from the {\tt Astropy} \citep{astropy} and {\tt Matplotlib} \citep{matplotlib} software packages.  This paper makes use of the following ALMA data: ADS/JAO.ALMA$\#$2015.1.00686.S, ADS/JAO.ALMA$\#$2011.1.00399.S, and ADS/JAO.ALMA$\#$2013.1.00198.S.  ALMA is a partnership of ESO (representing its member states), NSF (USA) and NINS (Japan), together with NRC (Canada), NSC and ASIAA (Taiwan), and KASI (Republic of Korea), in cooperation with the Republic of Chile.  The Joint ALMA Observatory is operated by ESO, AUI/NRAO, and NAOJ.  The National Radio Astronomy Observatory (NRAO) is a facility of the National Science Foundation operated under cooperative agreement by Associated Universities, Inc.  TB acknowledges support from the DFG through grant (KL 1469/13-1).

\end{document}